\documentclass[conference]{IEEEtran}
\IEEEoverridecommandlockouts

\usepackage{cite}
\usepackage{amsmath,amssymb,amsfonts}
\usepackage{algorithmic}
\usepackage{graphicx}
\usepackage{textcomp}

\usepackage{xcolor}

\usepackage{booktabs}
\usepackage{multirow}
\usepackage{array}
\usepackage{tikz}
\usepackage{titlesec}
\usepackage[none]{hyphenat}
\usepackage{float}
\def\BibTeX{{\rm B\kern-.05em{\sc i\kern-.025em b}\kern-.08em
    T\kern-.1667em\lower.7ex\hbox{E}\kern-.125emX}}

\begin{document}

\title{PAPR-Aware Multimodal Token Transmission in MLLM-Based Multiuser Networks}

\author{
\IEEEauthorblockN{Molka Trabelsi, Rafik Zayani}
\IEEEauthorblockA{
\textit{Univ Rennes, CNRS, IETR, UMR 6164, F-35000 Rennes, France}\\
\{molka.trabelsi, rafik.zayani\}@univ-rennes.fr}
}

\maketitle

\begin{abstract}
Task-oriented semantic communication (SemCom) empowered by multimodal large 
language models (MLLMs) has recently emerged as a promising paradigm for 
efficiently transmitting multimodal information, yet transmitting token 
embeddings over OFDM channels faces critical challenges due to hardware 
impairments, particularly the high peak-to-average power ratio (PAPR) that 
severely degrades energy efficiency under nonlinear power amplifiers.

In this paper, we propose a novel PAPR-aware task-oriented multimodal token transmission framework. Specifically, we propose MAPS (Multimodal AI-driven PAPR-aware Token Transmission) scheme for energy-efficient multiuser wireless communications. The key challenge is to jointly ensure inter-modal consistency, task-relevant token transmission, and PAPR reduction. To address this, we adopt a two-stage training strategy that integrates joint cross-modal alignment and PAPR reduction, followed by task-oriented fine-tuning.
MAPS transmits multimodal token embeddings over an OFDM channel under a realistic Modified Rapp power amplifier. Specifically, we incorporate a trainable linear projection in the text branch for effective gradient propagation, design a balanced multimodal PAPR loss with variance-based equalization across users, and anchor the reconstruction loss to a fixed semantic target to preserve cross-modal alignment.
Simulation results show that MAPS achieves balanced PAPR reduction across all modalities while maintaining strong inter-modal consistency. Under power amplifier nonlinearity, MAPS outperforms baseline schemes in both audio-visual question answering (AVQA) accuracy ($64.5\%$ gain) and task-oriented energy efficiency ($64.4\%$ gain) at an input back-off (IBO) of 3 dB, highlighting its effectiveness for energy-efficient multimodal semantic communication.
\end{abstract}

\begin{IEEEkeywords}
semantic communication, PAPR reduction, multimodal task-oriented communication, MLLMs, power amplifier, OFDM, 6G
\end{IEEEkeywords}


\section{Introduction}
\label{sec:intro}
 
Semantic communication has emerged as a key paradigm for 6G networks,
shifting the transmission objective from bit-level fidelity to
task-oriented meaning delivery~\cite{deepsc}.
Token-based communication (TokCom) is an emerging paradigm inspired
by recent advances in generative foundation models and Multimodal
Large Language Models (MLLMs), in which tokens serve as the fundamental
communication units carrying rich multimodal semantic information,
enabling efficient transformer-based processing at both the
transmitter and the receiver. Several works have explored this
direction: TokCom-MLMs~\cite{tokcom} proposes a multimodal token
transmission framework combining cross-modal alignment and
task-oriented fine-tuning; GenSemCom~\cite{gensemcom} develops
semantic-aware power allocation for generative MLMs; and
VQ-VAE~\cite{vqvae} explores discrete token representations with
importance-aware OFDM transmission. Critically, to the best of our knowledge, no existing work in the
open literature accounts for hardware impairments, and particularly
PA nonlinearity, in the context of multimodal token-based semantic
communication systems.
 
In this paper, we propose MAPS, a new PAPR-aware task-oriented multimodal 
token transmission framework for energy-efficient multiuser wireless 
communications over OFDM channels under a practical Modified Rapp power 
amplifier~\cite{rapp}. The contributions of this paper are summarized as 
follows:
\begin{figure*}[t]
\centering
\includegraphics[width=1\linewidth]{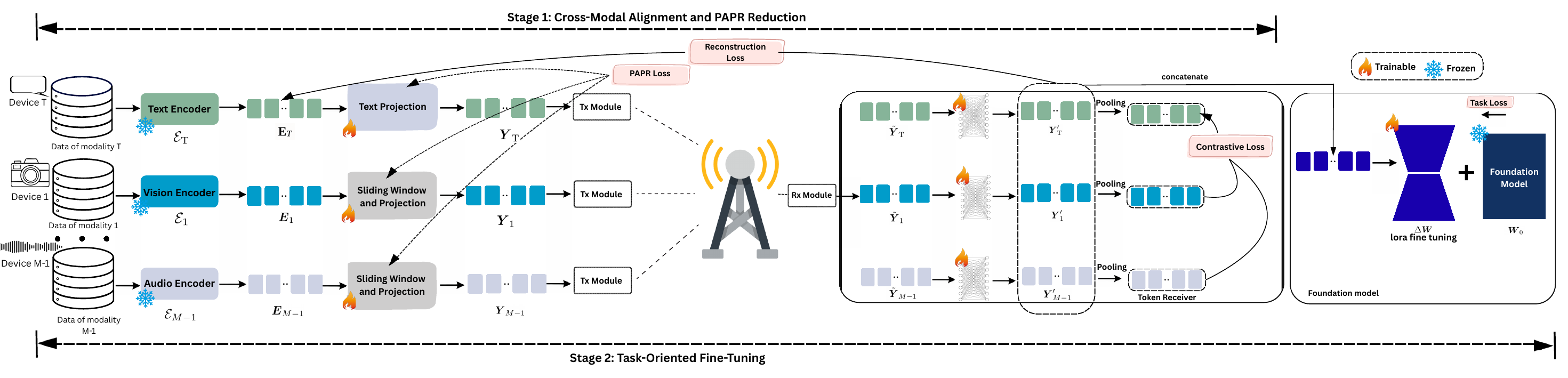}
    \caption{MAPS system architecture. Stage~1:
    joint PAPR reduction and cross-modal alignment. Stage~2: task-oriented fine-tuning via LoRA.
    Left: multiuser transmitter side with per-modality encoders,
    projections, and Tx modules. Right: BS receiver side with
    Rx modules, token receivers, and foundation model on the
    edge server.}
 \vspace{-13pt}   
\label{fig:arch}
\end{figure*}

\begin{itemize}
    \item To enhance the energy efficiency of task-oriented token
    transmission while preserving effective multimodal information
    exploitation, we propose a PAPR-aware two-stage training algorithm
    consisting of a joint PAPR reduction and cross-modal alignment
    stage, followed by task-oriented fine-tuning. A joint PAPR
    reduction and contrastive split fine-tuning strategy is first
    employed to project heterogeneous modalities into a unified
    feature space while simultaneously reducing the PAPR of OFDM
    signals across users. 

    \item Ensuring cross-modal alignment while simultaneously reducing
    the PAPR of the associated OFDM transmitted signals is highly
    non-trivial. In particular, introducing a PAPR reduction objective,
    especially in the presence of a trainable text projection, creates
    a moving reconstruction target that can destabilize cross-modal
    alignment. To address this challenge, we propose anchoring the
    reconstruction loss to a frozen embedding output, thereby decoupling
    PAPR optimization from the alignment process.

    \item Na\"{i}vely combining cross-modal alignment with PAPR reduction
    leads to uneven optimization across modalities, as stronger alignment
    gradients, particularly in the image branch, tend to resist PAPR
    reduction. To address this imbalance, we propose a balanced
    multimodal PAPR reduction strategy that incorporates an
    inter-modality variance regularization term that automatically
    equalizes gradient contributions across all modalities, ensuring
    uniform PAPR reduction while preserving effective cross-modal
    alignment.

    \item We introduce a task-oriented energy efficiency metric tailored
    to PAPR-aware multimodal token transmission. Numerical results show
    up to $65.5\%$ gain in task-oriented energy efficiency under PA
    nonlinearity and AWGN at low IBO, highlighting its effectiveness
    for energy-efficient multimodal semantic communication.
\end{itemize}

The rest of this paper is organized as follows: Section~\ref{sec:sysmodel}
presents the system model, Section~\ref{sec:method} describes the proposed method,
Section~\ref{sec:results} reports the results, and Section~\ref{sec:conclusion}
concludes the paper.

\noindent\textit{Notation.} Scalars $x$, vectors $\mathbf{x}$,
matrices $\mathbf{X}$.

\section{System Model}
\label{sec:sysmodel}

As shown in Fig.~\ref{fig:arch}, we consider a multiuser
wireless network consisting of $M$ user devices and a base station
(BS) equipped with an edge server. The network handles $M$
 data modalities, and each device processes one modality.
For notational simplicity, we define the set
$\mathcal{M} = \mathcal{M}' \cup \{T\}$ to index devices and
their corresponding modalities, where
$\mathcal{M}' = \{1, 2, \ldots, M-1\}$ denotes the non-text
modalities and $T$ denotes the text modality. Token
transmitters at the devices extract token embeddings from
raw data, modulate and transmit them over the wireless
channels. The received signal is processed at the BS by
a token receiver to recover token embeddings, which are then
concatenated and forwarded to the foundation model deployed
on the edge server for multimodal fusion and downstream
task execution.
 
\subsection{Transmitter}
 
\subsubsection{Token Transmitter}
 
For text data $\mathbf{X}_T$, the text encoder
$\mathcal{E}_T(\cdot)$ produces a token embedding sequence as
\begin{equation}
    \mathbf{E}_T = \mathcal{E}_T(\mathbf{X}_T),
    \quad \mathbf{E}_T \in \mathbb{R}^{s_T \times d_T},
    \label{eq:text_enc}
\end{equation}
where $s_T$ is the sequence length and $d_T$ is the token
dimension. For data $\mathbf{X}_m$ of non-text modality
$m \in \mathcal{M}'$, the corresponding encoder produces
\begin{equation}
    \mathbf{E}_m = \mathcal{E}_m(\mathbf{X}_m),
    \quad \mathbf{E}_m \in \mathbb{R}^{\hat{s}_m \times d_m},
    \label{eq:nontext_enc}
\end{equation}
where $\hat{s}_m$ and $d_m$ are the raw sequence length and
feature dimension. Non-textual encoders typically produce
sequences of excessively long length $\hat{s}_m$, which would
incur substantial bandwidth and latency overhead if transmitted
directly. To reduce the sequence length while retaining essential temporal patterns, 
we apply a sliding-window pooling approach: $\mathbf{E}_m$ is divided into 
$s_m$ consecutive token sets and a pooling operation compresses each set 
into a single token. Specifically, $S_m(\cdot)$ applies a non-overlapping 
sliding window of size $\lceil \hat{s}_m / s_m \rceil$ with mean pooling 
per window to produce a compressed sequence of length $s_m$. The compressed sequence is then mapped to
dimension $d_T$ via a fully connected layer $F_m(\cdot)$ giving
\begin{equation}
    \mathbf{Y}_m = F_m(S_m(\mathbf{E}_m)),
    \quad \forall m \in \mathcal{M}',
    \label{eq:nontext_proj}
\end{equation}
where $S_m(\cdot)$ is the sliding-window pooling module and
$\mathbf{Y}_m \in \mathbb{R}^{s_m \times d_T}$.
 
All encoders $\mathcal{E}_m(\cdot)$ are pretrained and kept
frozen throughout training. For the text branch, this freezing blocks gradient propagation from the PAPR loss back to the text embeddings. To address this, we introduce a trainable linear
projection $P_T(\cdot)$ between the frozen encoder output
and the OFDM modulator as
\begin{equation}
    \mathbf{Y}_T = P_T(\mathbf{E}_T),
    \label{eq:text_proj}
\end{equation}
where $\mathbf{Y}_T \in \mathbb{R}^{s_T \times d_T}$ are the
projected text token embeddings fed to the Tx module. The
design and training of this projection are detailed in
Section~\ref{sec:method}.
\hspace{-4mm}
\subsubsection{Tx Module : OFDM modulator and PA}
 
\begin{figure}[H]
    \centering
    \includegraphics[width=0.95\linewidth]{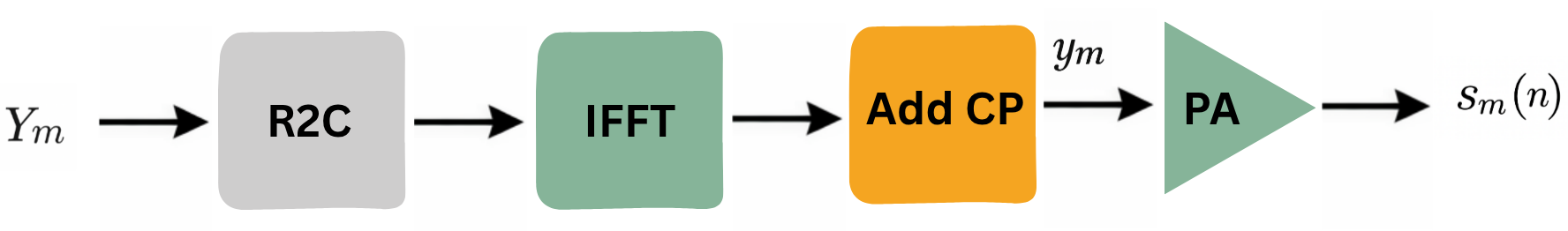}
    \caption{Per-modality Tx Module: token embeddings are
    converted to complex symbols (R2C), OFDM-modulated (IFFT),
    augmented with a cyclic prefix (Add CP), and passed through
    the power amplifier (PA) to produce the transmitted signal
    $s_m(n)$.}
    \label{fig:txmod}
\end{figure}
 
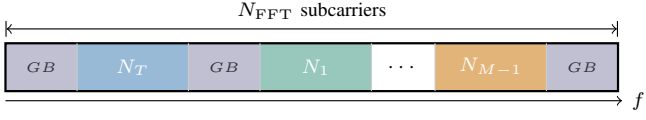
\begin{figure}[H]
\centering
\resizebox{1\linewidth}{!}{
\begin{tikzpicture}[font=\footnotesize\sffamily, scale=1, every node/.style={scale=1}]
  \definecolor{cText}{RGB}{70,130,180}
  \definecolor{cImg}{RGB}{95,170,150}
  \definecolor{cAud}{RGB}{210,140,50}
  \definecolor{cGB}{RGB}{150,150,180}

  \def\H{0.6}

  \def\wGB{0.9}
  \def\wNT{1.4}
  \def\wNOne{1.4}
  \def\wDots{0.8}
  \def\wNM{1.4}

  \def\xA{0}
  \def\xB{\xA+\wGB}
  \def\xC{\xB+\wNT}
  \def\xD{\xC+\wGB}
  \def\xE{\xD+\wNOne}
  \def\xF{\xE+\wDots}
  \def\xG{\xF+\wNM}
  \def\xH{\xG+\wGB}

  \fill[cGB!60]   (\xA,0) rectangle (\xB,\H);
  \fill[cText!55] (\xB,0) rectangle (\xC,\H);
  \fill[cGB!60]   (\xC,0) rectangle (\xD,\H);
  \fill[cImg!65]  (\xD,0) rectangle (\xE,\H);
  \fill[white]    (\xE,0) rectangle (\xF,\H);
  \fill[cAud!65]  (\xF,0) rectangle (\xG,\H);
  \fill[cGB!60]   (\xG,0) rectangle (\xH,\H);

  \draw[thick] (\xA,0) rectangle (\xH,\H);

  \foreach \x in {\xB,\xC,\xD,\xE,\xF,\xG} {
    \draw[thin,gray!40] (\x,0) -- (\x,\H);
  }

  \node[white,font=\scriptsize\bfseries] at ({(\xB+\xC)/2},\H/2) {$N_T$};
  \node[white,font=\scriptsize\bfseries] at ({(\xD+\xE)/2},\H/2) {$N_1$};
  \node[font=\scriptsize\bfseries]       at ({(\xE+\xF)/2},\H/2) {$\cdots$};
  \node[white,font=\scriptsize\bfseries] at ({(\xF+\xG)/2},\H/2) {$N_{M-1}$};

  \node[font=\tiny,text=cGB!30!black] at ({(\xA+\xB)/2},\H/2) {$GB$};
  \node[font=\tiny,text=cGB!30!black] at ({(\xC+\xD)/2},\H/2) {$GB$};
  \node[font=\tiny,text=cGB!30!black] at ({(\xG+\xH)/2},\H/2) {$GB$};

  \draw[|<->|,thin] (\xA,\H+0.18) -- (\xH,\H+0.18)
    node[midway,above,font=\scriptsize] {$N_{\mathrm{FFT}}$ subcarriers};

  \draw[->] (\xA,-0.12) -- (\xH+0.05,-0.12)
    node[right,font=\scriptsize] {$f$};

\end{tikzpicture}
}
\vspace{-0.75cm}
\caption{Subcarrier allocation: $N_{\mathrm{FFT}}$ subcarriers partitioned into modality-specific bands separated by guard bands.}
\label{fig:ofdm}
\end{figure}

As shown in Fig.~\ref{fig:txmod}, all token embeddings $\mathbf{Y}_m$ ($\forall m \in \mathcal{M}'$) are reshaped
and modulated to complex-valued signals, giving 
\begin{equation}
    \boldsymbol{y}_m = C_m(\mathbf{Y}_m), \quad \forall m \in \mathcal{M},
    \label{eq:mod}
\end{equation}
where $C_m(\cdot)$ performs R2C conversion, IFFT of size
$N_{\mathrm{FFT}}$, reshape and CP insertion of length
$N_{\mathrm{cp}}$. To avoid inter-modal interference, the
$N_{\mathrm{FFT}}$ subcarriers are partitioned into
$|\mathcal{M}|$ dedicated frequency bands as illustrated in
Fig.~\ref{fig:ofdm}, each of size $N_m$ subcarriers,
separated by guard bands of $N_g$ null subcarriers. This
orthogonal allocation ensures interference-free transmission
across modalities.
 
 \vspace{-0.25cm}
\subsubsection{Power Amplifier}
 
The PA implements AM/AM (amplitude-to-amplitude) and AM/PM
(amplitude-to-phase) distortions according to the Modified
Rapp model~\cite{rapp}. For input $y_m(n) = a_0(n)\,e^{j\theta(n)}$,
where $a_0(n)$ is the instantaneous amplitude and $\theta(n)$
is the phase and $n=0,...,N_{FFT}-1$. The PA output per modality is
\begin{equation}
    s_m(n) = \mathrm{PA}(y_m(n)) =
    A(\alpha a_0(n))\,e^{j(\theta(n)+\phi(\alpha a_0(n)))}.
    \label{eq:pa}
\end{equation}
The factor $\alpha$ is a multiplicative scaling coefficient
applied at the PA input to operate at a specified IBO level.
In this work, all IBO values are expressed in dB relative to
the back-off from the saturation point, defined as the
operating point where the output power reaches saturation in
an equivalent linear amplifier. To ensure that the transmitted
signal $y_m(n)$ meets a target IBO level, the scaling factor
is given by:
\begin{equation}
    \alpha = \sqrt{\frac{P_{\mathrm{sat}}}{10^{\mathrm{IBO}/10}\,P_t}},
    \label{eq:alpha}
\end{equation}
where $P_{\mathrm{sat}}$ is the input saturation power and $P_t$ is the average input power. The IBO is
defined as
$\mathrm{IBO} = 10\log_{10}(P_{\mathrm{sat}}/P_{\mathrm{in}})$\,dB
and is swept at evaluation. The AM/AM and AM/PM \cite{rapp} responses are 
\begin{equation}
    A(a_0(n)) = \frac{G\,a_0(n)}{\left(1+\left|\frac{G\,a_0(n)}
        {A_{\mathrm{sat}}}\right|^{2p}\right)^{1/(2p)}},
    \label{eq:amam}
\end{equation}
\begin{equation}
    \phi(a_0(n)) = \frac{A_\phi\,a_0(n)^q}{1+(a_0(n)/B_\phi)^q}.
    \label{eq:ampm}
\end{equation}
where $G$ is the small-signal gain, $A_{\mathrm{sat}}$ is the
saturation amplitude, $p$ is the AM/AM smoothness parameter,
and $A_\phi$, $B_\phi$, $q$ are the AM/PM parameters.
During training, the PA is disabled, and the composite
received signal at the BS via the wireless channel is given by
\begin{equation}
    r(n) = \sum_{m \in \mathcal{M}} h_ms_m(n) + w(n), \forall n=0,...,N_{FFT}-1,
    \label{eq:composite_rx}
\end{equation}
where $h_m$ is the channel coefficient and $w(n)$ is the additive noise 
sample at time index $n$, which follows a circularly symmetric complex 
Gaussian distribution $\mathcal{CN}(0, \sigma^2)$, with zero mean and 
variance $\sigma^2$.
 
\subsection{Receiver}
 
\subsubsection{Rx module: OFDM Demodulation}
 
\begin{figure}[H]
    \centering
    \includegraphics[width=0.75\linewidth]{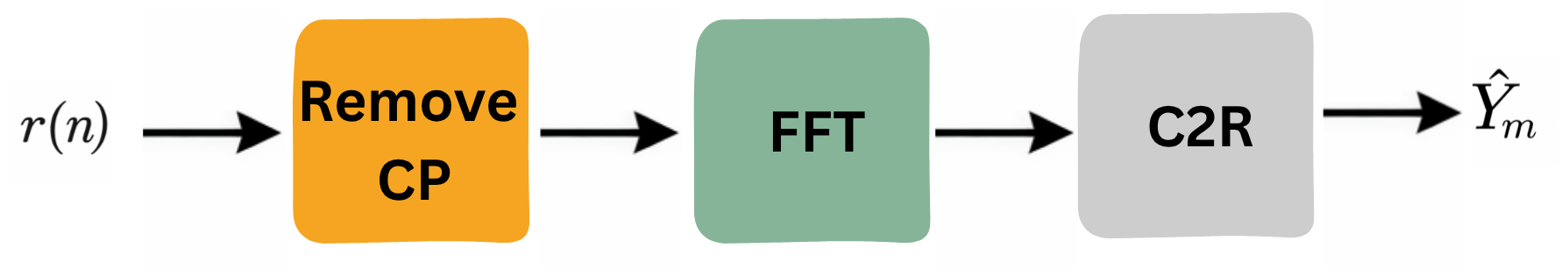}
    \caption{Rx Module: the received signal
    $\boldsymbol{r}$ is processed by CP removal, FFT, C2R and reshaped to recover the modality-specific received signals $\tilde{\mathbf{Y}}_m \forall m \in  \mathcal{M}$.}
    \label{fig:rxmod}
\end{figure}
  \vspace{-0.25cm}
At the BS (Fig. \ref{fig:rxmod}), the received signal $\boldsymbol{r}$ is demodulated and reshaped back into token space giving each modality-specific tokens as
\begin{equation}
    \tilde{\mathbf{Y}}_m = C_m^{-1}(\boldsymbol{r}),
    \quad \forall m \in \mathcal{M},
    \label{eq:demod}
\end{equation}
where $\tilde{\mathbf{Y}}_m \in \mathbb{R}^{s_m \times d_T} $ and $C_m^{-1}(\cdot)$ performs CP removal, FFT, C2R and reshape as shown in Fig.~\ref{fig:rxmod}.
 
\subsubsection{Token Receiver}
 
A two-layer MLP token receiver $G_m(\cdot)$ with ReLU
activation refines the demodulated embeddings
\begin{equation}
    \mathbf{Y}'_m = G_m(\tilde{\mathbf{Y}}_m),
    \quad \forall m \in \mathcal{M},
    \label{eq:receiver}
\end{equation}
where $\mathbf{Y}'_m \in \mathbb{R}^{s_m \times d_T}$.

The recovered embeddings from all modalities are concatenated
and forwarded to the foundation model, which gives the task output as
\begin{equation}
O = \mathcal{W}\bigl(
        U({\mathbf{Y}'}_1, \ldots,
        {\mathbf{Y}'}_{M-1}, {\mathbf{Y}'}_T)
    \bigr),
    \label{eq:fusion}
\end{equation}
where $U(\cdot)$ is the concatenation module,
$\mathcal{W}(\cdot)$ is the foundation model on the edge
server, and $O$ is the task output. 
\subsection{Performance Evaluation Metrics}

We evaluate system performance using three complementary
metrics. The PAPR of modality $m$ is defined as
\begin{equation}
    \mathrm{PAPR}_m =
    \frac{\max_n |y_m(n)|^2}{\mathbb{E}[|y_m(n)|^2]},
    \label{eq:papr}
\end{equation}
expressed in dB as $10\log_{10}(\mathrm{PAPR}_m)$ where $y_m(n)$ denotes the $n$-th entry of $\boldsymbol{y}_m$ 
in~\eqref{eq:mod}, $\forall m \in \mathcal{M}$. The
complementary cumulative distribution function (CCDF) of
the PAPR, defined as $\mathrm{Pr}(\mathrm{PAPR}_m > \gamma_0)$,
characterizes the probability that the instantaneous PAPR
exceeds a given threshold $\gamma_0$, and is used to evaluate
PAPR reduction across modalities.

Task performance is assessed via audio-visual question
answering (AVQA) accuracy $A(\mathrm{IBO})$, measured on
the test set under PA nonlinearity at a given IBO level. For a predicted word $\hat{o}_i$ and the corresponding ground-truth word $o_i$, the accuracy is computed as
\begin{equation}
    A(\mathrm{IBO}) = \frac{1}{N} \sum_{i=1}^{N} \mathbf{1}\{\hat{o}_i = o_i\},
\end{equation}
where $N$ denotes the total number of test samples and $\mathbf{1}\{\cdot\}$ be the indicator function.

The task-oriented energy efficiency is defined as
\begin{equation}
    \mathrm{EE}_{\mathrm{TO}} =
    \frac{D_{\mathrm{avg}} \cdot A(\mathrm{IBO})}
         {P_{\mathrm{tot}}(\mathrm{IBO})},
    \label{eq:ee}
\end{equation}
where $D_{\mathrm{avg}}$ is the average query throughput in queries 
per second and $P_{tot}$ is the total power consumed, which can be defined as 
\begin{equation}
    P_{\mathrm{tot}} =
    \frac{A_{\mathrm{sat}}^2\sqrt{10^{\mathrm{IBO}/10}}}
         {\eta_{\max}}
    + P_0 + \varepsilon R,
    \label{eq:ptot}
\end{equation}
where $\eta_{\max}$ is the maximum PA efficiency, so that 
$A_{\mathrm{sat}}^2\sqrt{10^{\mathrm{IBO}/10}}/\eta_{\max}$ 
represents the PA power consumption at a given IBO level, 
$P_0$ is the baseband and RF circuit power, $\varepsilon$ is the
processing cost per bit, and $R$ is the sum data rate, so that 
$\varepsilon R$ accounts for the digital processing power.
\section{Two-stage training based PAPR-aware multimodal task-oriented  token transmission}
\label{sec:method}
 
\subsection{Training Overview}

MAPS follows the two-stage training protocol of~\cite{tokcom},
extended to incorporate hardware impairment awareness through
PAPR-aware optimization.

In \textbf{Stage~1}, the system jointly learns cross-modal alignment
and waveform shaping under physical constraints. The goal is to map
all modalities into a shared semantic space while simultaneously
reducing the PAPR of the OFDM modulated signals.

In \textbf{Stage~2}, the aligned representations are fine-tuned for
the downstream AVQA task. The two stages are decoupled to preserve
semantic consistency while enabling efficient task adaptation.

The PA is disabled during training and applied
only at inference time.

\subsection{Stage~1: Joint PAPR Reduction and Cross-Modal Alignment}

Stage~1 optimizes semantic alignment and PAPR reduction jointly.
Cross-modal alignment ensures that token embeddings from different
modalities lie in a shared feature space, while PAPR reduction
shapes the modulated signals to be hardware friendly.

Alignment is enforced through a combination of reconstruction and
contrastive objectives. For each modality $m \in \mathcal{M}'$, a
pooled representation $a_m$ is extracted and aligned with the text
representation $a_T$ using
\begin{equation}
\mathcal{L}_{\mathrm{CON}}^m =
- \log \frac{\exp(\mathrm{sim}(a_m, a_T)/\tau)}
{\sum_{k=1}^{K} \exp(\mathrm{sim}(a_k, a_T)/\tau)},
\quad \forall m \in \mathcal{M}',
\end{equation}
where $\mathrm{sim}(\cdot,\cdot)$ denotes cosine similarity, $\tau$ is 
a temperature parameter, and $K$ is the batch size.

To enable PAPR optimization on the text modality, we introduce a
trainable projection between the frozen text encoder output and the OFDM modulator.
This projection provides a gradient path while preserving the
semantic structure of the embeddings.

To ensure consistent PAPR reduction across modalities, we define a
balanced multimodal PAPR loss
\begin{equation}
\mathcal{L}_{\mathrm{PAPR}} =
    \alpha_1\,\bar{\rho} + \alpha_2\,\rho_{\max}
    + \alpha_3\,\mathrm{Var}(\rho_m)
    + \alpha_4\,\rho_{\mathrm{img}},
\label{eq:papr_loss}
\end{equation}
where $\alpha_1, \alpha_2, \alpha_3, \alpha_4 \geq 0$ are non-negative 
weighting coefficients controlling the relative importance of the mean 
PAPR, worst-case PAPR, inter-modality variance, and image-branch pressure 
terms, respectively, $\rho_m = \max_n |y_m(n)|^2 / \mathbb{E}[|y_m(n)|^2]$
denotes the PAPR of modality $m$ in linear scale,
$\bar{\rho} = \frac{1}{|\mathcal{M}|}\sum_{m \in \mathcal{M}}\rho_m$
is the mean PAPR across modalities driving global reduction,
$\rho_{\max} = \max_{m \in \mathcal{M}} \rho_m$ targets the
worst-case modality at each step,
$\mathrm{Var}(\rho_m) = \mathbb{E}[(\rho_m - \bar{\rho})^2]$
penalizes inter-modality imbalance by automatically increasing
gradient pressure on lagging branches, and $\rho_{\mathrm{img}}$
applies dedicated constant pressure on the image branch to
compensate for its strong alignment gradients that would
otherwise resist PAPR reduction.

Therefore, the overall Stage-1 loss is
\begin{equation}
    \mathcal{L} =
    \mathcal{L}_{\mathrm{MSE}}
    + \mathcal{L}_{\mathrm{CON}}
    + \lambda\,\mathcal{L}_{\mathrm{PAPR}},
\label{eq:loss}
\end{equation}
where normalization ensures stable joint optimization.

Finally, to stabilize alignment, the reconstruction loss is anchored
to the text encoder output:
\begin{equation}
\mathcal{L}_{\mathrm{MSE}} =
\bigl\|\mathbf{E}_{\text{text}}^{\mathrm{raw}}
- \hat{\mathbf{E}}_{\text{text,rx}}\bigr\|^2
\label{eq:anchor}
\end{equation}
where $\mathbf{E}_{\text{text}}^{\mathrm{raw}}$ remains fixed during
training. This prevents the alignment target from drifting and
ensures stable optimization despite the presence of trainable
projection layers.

During this stage, only the projection layers are trained, while
the modality encoders and the foundation model remain frozen.

\subsection{Stage~2: Task-Oriented Fine-Tuning}

After cross-modal alignment, the system is fine-tuned for the
downstream AVQA task. The received token embeddings are
concatenated and forwarded to the foundation model deployed at the edge server, as given in (\ref{eq:fusion}).

We adopt parameter-efficient fine-tuning using LoRA~\cite{lora}.
Given a pretrained model $\mathbf{W}_0$, we introduce a low-rank adaptation
\begin{equation}
    \Delta \mathbf{W} = \mathbf{B}\mathbf{A}^\top,
\end{equation}
yielding the adapted model
\begin{equation}
    \mathbf{W} = \mathbf{W}_0 + \Delta \mathbf{W}.
\end{equation}

The system is optimized using a task-specific loss
\begin{equation}
    \mathcal{L}_{\mathrm{TSK}} =
    \ell(\hat{\mathbf{Y}}; \mathbf{W}_0, \Delta \mathbf{W}),
\end{equation}
where $\ell(\cdot)$ denotes the cross-entropy loss.

During Stage~2, all Stage-1 components are frozen, and only the
LoRA parameters are updated, preserving the learned semantic
alignment while enabling efficient task adaptation.
\section{Simulations Results}
\label{sec:results}
 
\subsection{Scenario Setup}
In the simulations, we consider a multimodal task-oriented
token transmission framework for MLLM-based multiuser
communication involving textual, audio, and visual modalities.
All user devices transmit with power $P_t$, which is
directly related to the modality-specific IBO. The foundation model used is
Qwen2.5-1.5B~\cite{qwen} equipped with a low-rank adaptation
(LoRA) module of rank $r=4$, deployed at the edge server.

For modality-specific encoding, we adopt the Qwen2.5 text
encoder for textual data, ViViT-B/16$\times$2~\cite{vivit}
for visual inputs, and the Audio Spectrogram Transformer
(AST)~\cite{ast} fine-tuned on AudioSet for audio. All
encoders are frozen throughout training. The trainable
components are the MAPS-Text projection, the image and audio
transmitter projections, and the token receivers. It is worth
noting that our focus is on the joint design of PAPR reduction
and cross-modal alignment. Therefore, we adopt the power
allocation strategy from~\cite{tokcom}, which ensures efficient
token transmission while allowing us to isolate and evaluate
the impact of the proposed framework.

Stage~1 is trained on the VALOR dataset~\cite{valor} for
joint cross-modal alignment and PAPR reduction. Stage~2 is
fine-tuned on the MUSIC-AVQA dataset~\cite{musicavqa} for
the AVQA task, and evaluation is performed on the MUSIC-AVQA
test set. All results are reported under PA nonlinearity at
SNR\,=\,12\,dB and AWGN channel unless otherwise stated. The baseline is a
Stage-2 fine-tuned model without PAPR-aware training, using
the same channel and PA configuration. System parameters are
listed in Table~\ref{tab:params}.
 
\begin{table}[H]
\caption{System Parameters and Training Configurations}
\label{tab:params}
\centering
\footnotesize
\renewcommand{\arraystretch}{0.9}
\setlength{\tabcolsep}{3pt}
\begin{tabular}{lll}
\toprule
\textbf{Parameter} & \textbf{Symbol} & \textbf{Value} \\
\midrule
\multicolumn{3}{l}{\textit{OFDM modulation}} \\
FFT size / BW & $N_{\mathrm{FFT}}/B_m$ & $2448 / N_m \times 15\text{ kHz}$ \\
Guard band / CP & $N_g/\rho$ & $3/7\%$ \\
Training SNR & $\gamma$ & $12\text{ dB}$ \\
\midrule
\multicolumn{3}{l}{\textit{PA Model (Modified Rapp)~\cite{rapp}}} \\
$A_{\mathrm{sat}}/G/p/q$ & --- & $1.9/16/1.1/4.0$ \\
Max PA efficiency & $\eta_{\max}$ & $0.785$ \\
\midrule
\multicolumn{3}{l}{\textit{Model \& Training}} \\
Token dim / Text / Image\&Audio & $d_T/s_T/s_m$ & $1536/32/128$ \\
Stage-1 / Stage-2 iter & --- & $1000/300$ \\
Batch / LR & --- & $32/10^{-4}$ \\
Temp. / PAPR weight & $\tau/\lambda$ & $0.07/5$ \\
\bottomrule
\end{tabular}
\end{table}
 
\subsection{Stage-1 Training Convergence}
\vspace{-0.35cm}
\begin{figure}[H]
    \centering
    \includegraphics[width=\linewidth]{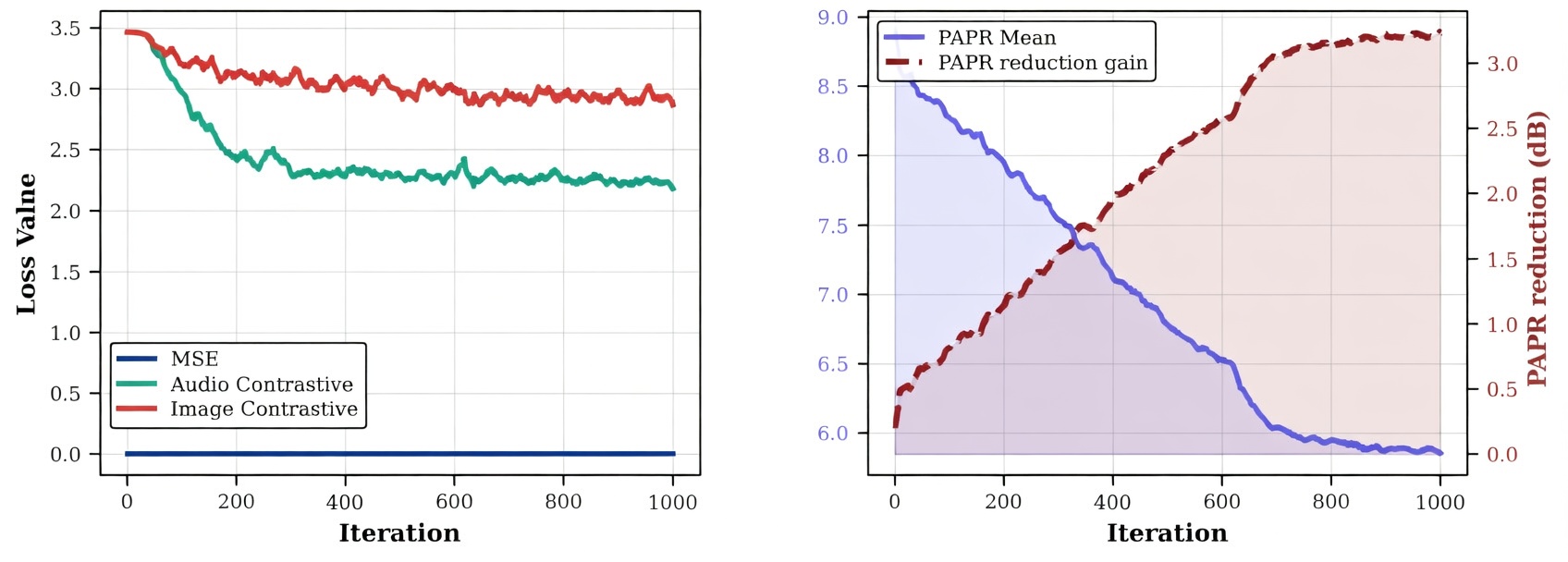}
    \vspace{-0.75cm}
    \caption{Stage-1 training curves. Left: alignment loss
    components --- MSE~\eqref{eq:anchor}, audio and image
    contrastive losses from~\eqref{eq:loss}. Right: mean
    PAPR evolution and cumulative reduction of
    $\mathcal{L}_{\mathrm{PAPR}}$~\eqref{eq:papr_loss}.}
    \label{fig:training}
\end{figure}
Fig.~\ref{fig:training} shows the Stage-1 training dynamics.
The alignment losses converge smoothly throughout training:
the audio contrastive loss decreases from $3.47$ to $2.22$
and the image contrastive loss from $3.47$ to $2.83$, while
the MSE loss remains near zero, confirming that the fixed
anchor~\eqref{eq:anchor} keeps the reconstruction target stable.
Simultaneously, the mean PAPR across modalities drops from $9.09$\,dB to
approximately $6.0$\,dB, with the cumulative reduction
reaching $3.1$\,dB by iteration~1{,}000. These curves confirm
that PAPR reduction and cross-modal alignment converge
simultaneously without interfering with each other.
 
\subsection{PAPR Reduction}

\begin{figure}[H]
    \centering
    \includegraphics[width=0.85\linewidth]{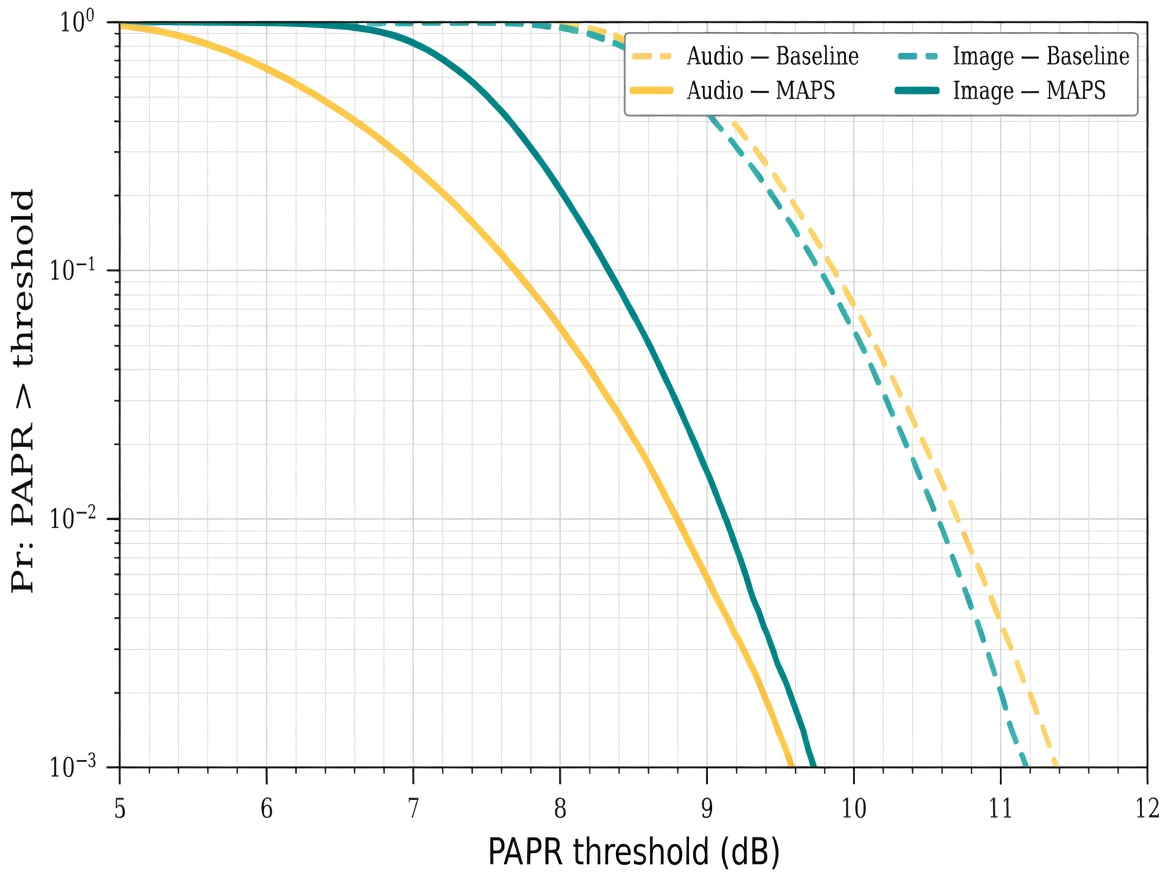}
    \caption{Stage 2 PAPR reduction across modalities (Image and Audio).}
    \label{fig:ccdf}
\end{figure}

Fig.~\ref{fig:ccdf} presents the PAPR complementary
cumulative distribution function (CCDF) measured on the
test set after Stage-2 training. MAPS achieves a mean PAPR reduction of 2.62 dB and 1.80 dB for, respectively, audio and image-related OFDM signals, at CCDF of $10^{-2}$.
 
\subsection{Accuracy under PA Nonlinearity}
\vspace{-0.35cm}
\begin{figure}[H]
    \centering
    \includegraphics[width=0.85\linewidth]{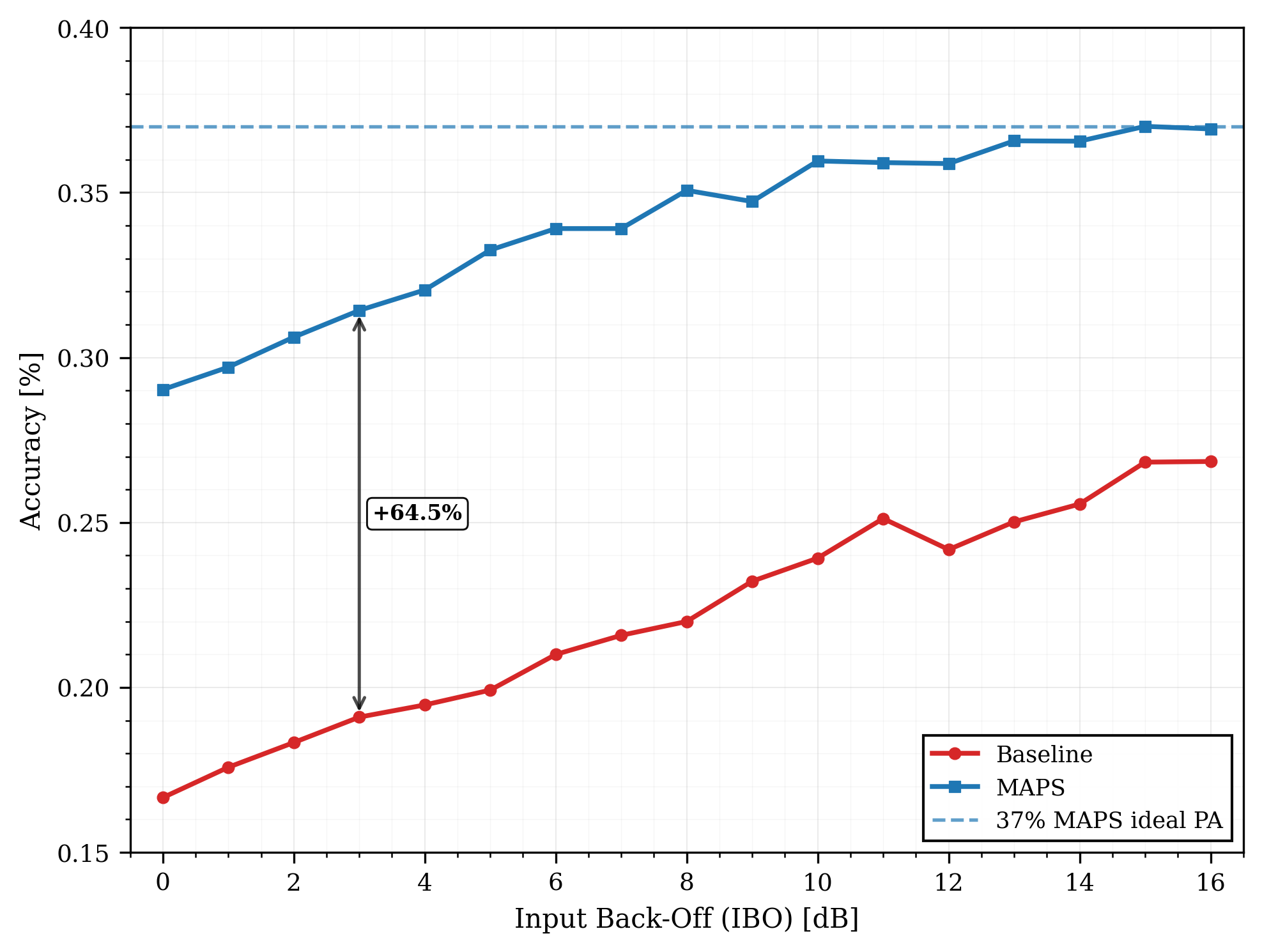}
    \caption{AVQA accuracy vs.\ IBO (SNR\,=\,12\,dB,
    measured after Stage-2).}
    \label{fig:ibo}
\end{figure}

Fig.~\ref{fig:ibo} presents AVQA accuracy as a function
of IBO from 1 to 16\,dB. MAPS outperforms the baseline
at every tested IBO value. The gain reaches $64.5\%$
at IBO\,=\,3\,dB, where the PA operates near the saturation, leading to high power efficiency and then high global system energy efficiency. The gain decreases as IBO increases and the MAPS performance approaches the baseline with ideal PA, but remains positive
across the full range, confirming that PAPR-aware
Stage-1 training translates directly into improved task
accuracy under realistic PA operating conditions.
 
\subsection{SNR Robustness}
 \vspace{-0.35cm}

\begin{figure}[H]
    \centering
    \includegraphics[width=0.85\linewidth]{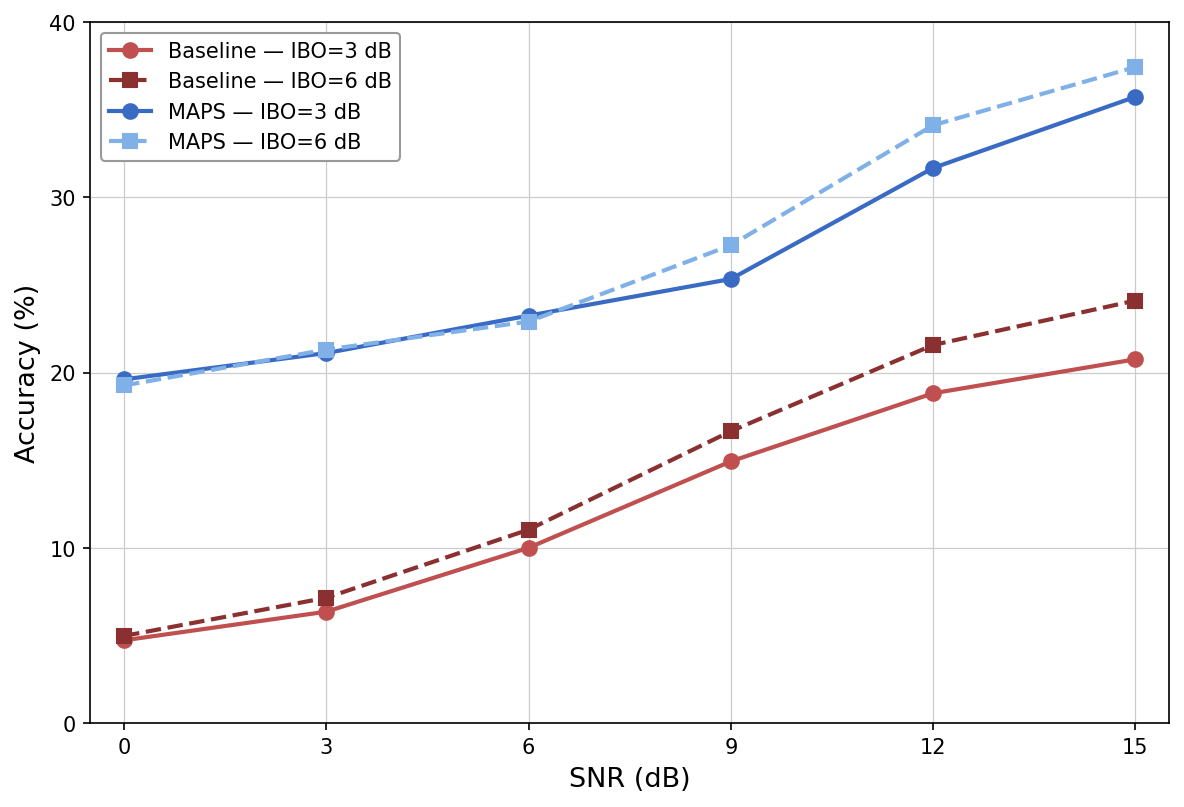}
    \caption{AVQA accuracy vs.\ SNR.}
    \label{fig:snr}
\end{figure}
Fig.~\ref{fig:snr} reports accuracy versus SNR at fixed
IBO\,=\,3\,dB and IBO\,=\,6\,dB. MAPS consistently
outperforms the baseline across the full SNR range from
0 to 15\,dB at both operating points. At the training
SNR of 12\,dB, MAPS achieves
$31.7\%$ at IBO\,=\,3\,dB versus $18.8\%$ for the
baseline, a gain of 68.31\%. The consistent advantage
across all SNR values confirms that the benefits of
PAPR-aware training are not limited to a specific channel
condition.
 
\subsection{Task-Oriented Energy Efficiency}
\vspace{-0.35cm}
\begin{figure}[H]
    \centering
    \includegraphics[width=0.85\linewidth]{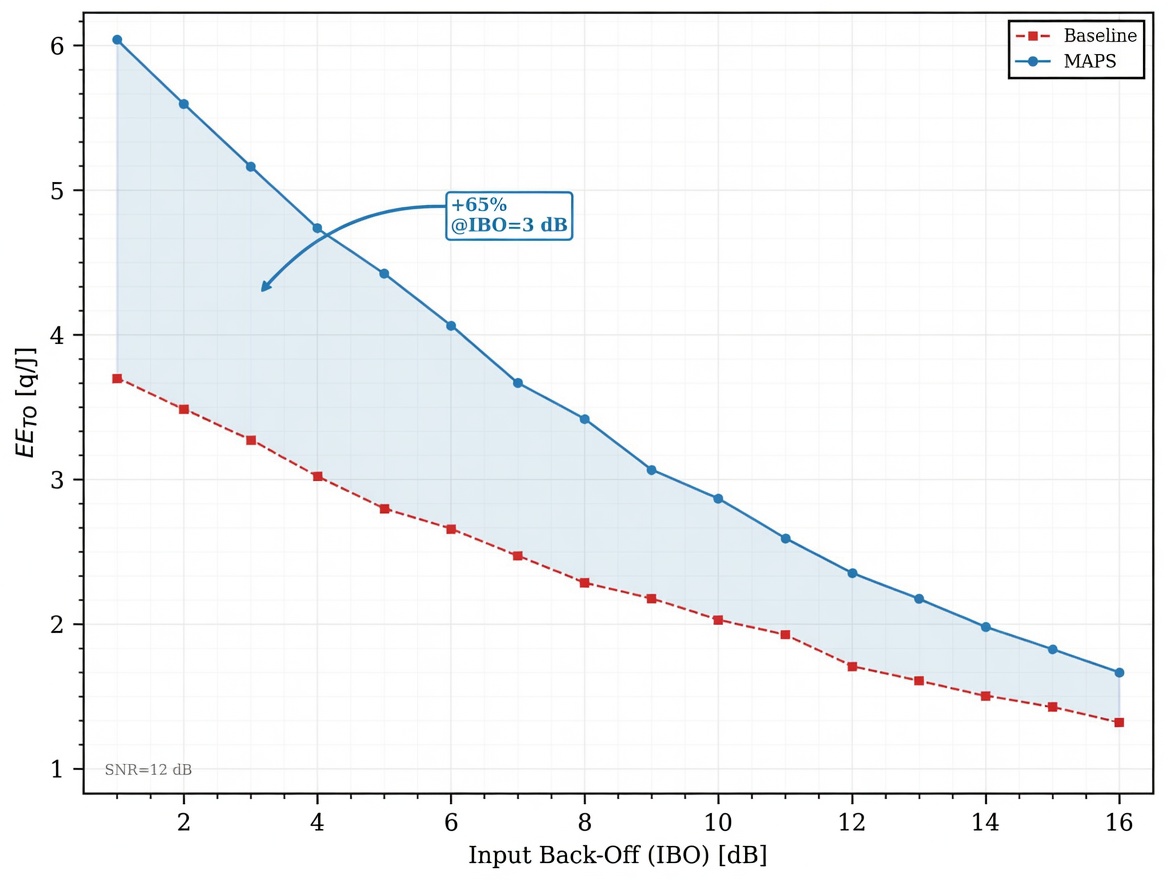}
    \caption{Task-oriented energy efficiency
    $\mathrm{EE}_{\mathrm{TO}}$ vs.\ IBO
    (SNR\,=\,12\,dB, measured after Stage-2).}
    \label{fig:ee}
\end{figure}
Fig.~\ref{fig:ee} reports the task-oriented energy
efficiency $\mathrm{EE}_{\mathrm{TO}}$ defined
in~\eqref{eq:ee}. MAPS achieves higher energy efficiency
than the baseline at every tested IBO value. The gain is
largest at low IBO, where both accuracy improvement and
power savings combine: at IBO\,=\,3\,dB, MAPS achieves
$5.46$\,q/J versus $3.32$\,q/J for the baseline, a gain
of $64.4\%$. As IBO increases and the PA becomes more
linear, the gap narrows but remains consistently positive,
reaching $37.5\%$ at IBO\,=\,16\,dB. Simulation results
demonstrate that PAPR-aware training improves task-oriented
energy efficiency across the full range of PA operating
points, making MAPS well-suited for deployment in
energy-constrained wireless systems.


\section{Conclusion}
\label{sec:conclusion}

This paper presented MAPS, a PAPR-aware multimodal token transmission framework that extends TokCom-MLMs to account for per-modality power amplifier nonlinearity. The proposed two-stage training strategy jointly optimizes PAPR reduction and cross-modal alignment: (1) an inter-modality variance regularization term automatically equalizes gradient contributions across modalities, while anchoring the reconstruction loss to a frozen embedding output decouples waveform shaping from semantic alignment. (2) Simulation results on MUSIC-AVQA demonstrate PAPR reductions of $2$\,dB and $1.8$\,dB for audio- and image-related OFDM signals at a CCDF of $10^{-2}$, a $64.5\%$ accuracy gain, and about $64.4\%$ improvement in task-oriented energy efficiency at IBO\,=\,3\,dB. By shifting PAPR control from circuit design to the learned embedding space, MAPS opens a path toward practical wireless transmission of semantic representations where energy efficiency and semantic fidelity are jointly optimized. Future works will explore advanced resource allocation strategies associated with multi-path and multiple-input multiple-output (MIMO) fading channels, asynchronism between users, and distributed edge AI servers orchestration.
\vspace{-0.15cm}
\section*{Acknowledgment}
This work was supported by the French State via the French National Research Agency (ANR) under the France 2030 program: RIS3 (ANR-23-CMAS-0023) and NEXCOM (ANR-25-PEFT-0004).

\end{document}